**The Formation of Ultra-Stable Glasses via Precipitation: a Modelling Study**

Ian Douglass and Peter Harrowell

School of Chemistry, University of Sydney, Sydney 2006 NSW, Australia

Abstract

The precipitation of a glass forming solute from solution is modelled using a lattice model previously introduced to study dissolution kinetics of amorphous materials. The model includes the enhancement of kinetics at the surface of a glass in contact with a plasticizing solvent. We demonstrate that precipitation can produce a glass substantially more stable than that produced by very long time annealing of the bulk glass former. The energy of these ultra-stable amorphous precipitates is found to be dominated by residual solvent rather than high energy glass configurations.

Glass films formed by vapor deposition can exhibit enthalpies and volumes significantly lower than the analogous glass formed by cooling of the bulk liquid [1-3]. In the case of vapor deposited indomethacin, for example, the *equilibrium* density was obtained in a film 25°C below the glass transition temperature, a result that would have required somewhere between $10^2$ and $10^4$ years aging of a bulk sample [4]. Very stable amorphous films offer a range of attractive properties that include chemical stability [5] and the selection of molecular orientation [6] with potential applications that include organic electronics [7].

Vapor deposition, while clearly representing an important path to the fabrication of novel amorphous materials, is not without its drawbacks. The vacuum chamber and pumps represent a considerable expense and the deposition requires the vaporization of the material to be deposited, a problem for molecules with low vapor pressures or thermally unstable species. In this paper we examine the possibility of avoiding these problems by forming an ultra-stable glass by precipitation from solution at or below room temperature. Precipitation can take place either via homogeneous nucleation in the bulk of a solution or heterogeneously on a substrate surface. While we focus on the former process in this paper, the main conclusions also apply to substrate deposition as demonstrated in the Supplementary Material.

The essential physics of the problem can be understood from the schematic phase diagram in Fig. 1. The equilibrium phase behaviour is provided by the solid-liquid coexistence curve where x is the mole fraction of solute. This curve establishes the temperature dependence of the saturation concentration and the melting point of the pure solute. In the context of glass formers, we shall assume that the crystal is kinetically inaccessible so that we can neglect the equilibrium diagram in favour of the phase behaviour of the metastable solution. That leaves us with the (metastable) liquid-liquid coexistence, characterized by a critical temperature $T_c$, characterising the thermodynamics that will drive precipitation. Here we assume that the solvent crystallization occurs at a temperature well below the glass transition of the solute and can therefore be neglected. To this phase diagram we add the (non-equilibrium) glass transition line which describes how the glass transition temperature $T_g$ of the solute depends on the solvent concentration. The composition dependence of $T_g$ in a binary mixture has been studied extensively [8-11]. In most cases the variation can be modelled either as a linear interpolation between the $T_g$'s of the two pure species [8] or as a modest nonlinear variant on the linear expression [11]. In this context, the low $T_g$ component is often referred to as a *plasticizer*. The glass transition line will cross the binodal line at some of value of $T = T^*$, the value of which depends on the relative values of $T_c$ and $T_g$. As is evident from the diagram in Fig. 1, precipitation at temperatures below $T^*$ will result in a solute rich glass, rather than a liquid, with a composition determined by $T_g(x)$ rather than the (metastable) equilibrium value.

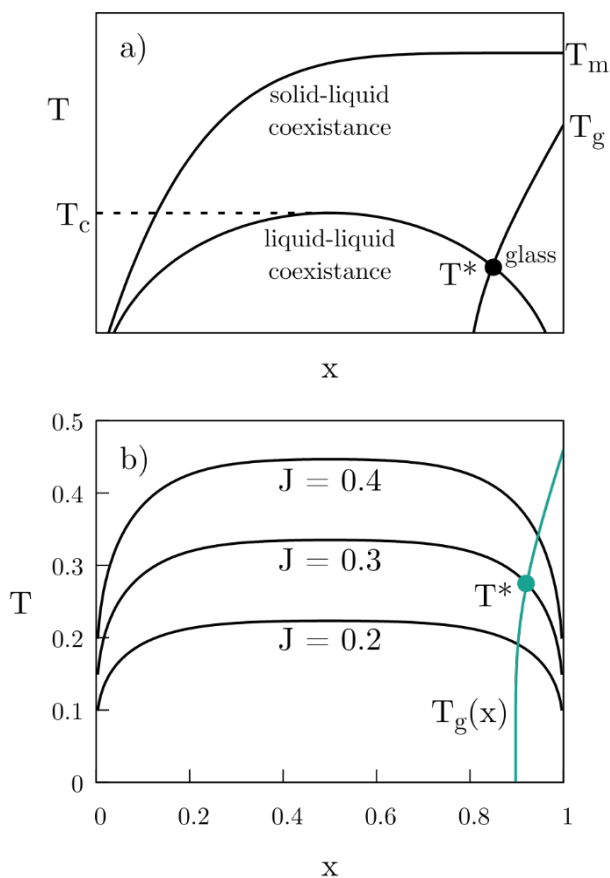

**Figure 1.** a) A schematic phase diagram for a mixture of solvent and solute with x being the mole fraction of solute and the concentration dependence of the solute glass transition temperature $T_g$. Also included is $T^*$ as defined in the text. b) The calculated liquid-liquid phase diagram for three different values of J, the solvent-solute mixing energy. The glass transition line $T_g(x)$ is also included. For J = 0.3, $T^* = 0.275$.

The formation of amorphous solids during phase separation has a considerable literature and we shall try, here, to briefly review these studies. It is well established [12] that some inorganic salts precipitate from solution into an initial amorphous solid before crystallizing, post-precipitation, from the solid phase. These amorphous intermediates play a significant role in biomineralization [13]. The precipitation of silica from solution typically involves the formation of a gel-like solid prior to complete crystallization [14]. The amorphous precipitation during de-mixing represents a general route to gel formation and this process has been described with the same generic diagram shown in Fig. 1 [15]. Glass formation in liquids characterised by short range attractions can pre-empted by the arrest of a low density aggregate during precipitation [16]. The coincidence of both processes – gelation and vitrification – has been reported [17]. The analogous processes of condensation of a glass from a vapor [18] and from colloidal suspensions [19] have also been studied. There has also been considerable research into the phase separation of kinetically asymmetric liquids, i.e. liquids characterised by a large difference in their respective $T_g$'s [20]. While the literature summarised here includes a wealth of information about the (non-equilibrium) phase diagrams and morphology of amorphous precipitation, we are unware of any previous study of the subject of this paper, i.e. the stability of the resulting amorphous materials and, specifically, the conditions under which this stability might be optimised.

The essential feature of glass physics responsible for the increased stability of glasses formed by vapor deposition is the enhanced kinetics at the glass surface [21] and, hence, it is necessary that our model properly captures this enhanced surface kinetics. We note that this enhancement is a feature of the thermally equilibrated surface [21] and so is not simply a consequence of a high initial kinetic energy of deposited particles. The premise of the model is that a supercooled liquid or glass is characterised by fluctuations in structure that strongly influence the local kinetics. If we imagine that we can capture that aspect of structure that exerts this kinetic influence – let's call this quantity σ – then we do not need to distinguish all the different possible structures. It is assumed that the kinetically inert domains correspond to low energy states. The facilitated kinetic Ising model is a simple expression of this physical situation where the quantity σ can take on just two values locally – a high energy value (σ = 1 or 'spin up') and a low energy value (σ = 0 or 'spin down') with an energy difference $h$ between the two states. While the model does make use of periodic lattice, the essential fluctuations (i.e. those of the spin variables) are highly disordered. The coupling between structure and dynamics is introduced as an explicit expression of the spin flip probability in terms of the spin states of the nearest neighbours. Specifically, a spin cannot flip, up or down, unless it has (on a simple cubic lattice) at least 3 up spins on neighbouring particles. This condition becomes increasingly harder to satisfy as the concentration of up spins decreases on cooling. The resulting dynamics – non-Arrhenius and spatial heterogeneous – provides a physically reasonable account of dynamics in a glass forming liquid [22].

The kinetic Ising model has been successfully extended to modelling vapour deposition [23] and, more recently, the kinetics of amorphous dissolution [24]. For this latter problem, a second component – the solvent – was introduced. The solvent particle interacts with each neighbouring solute particle with an energy J. We shall consider the case of a positive heat of mixing, i.e. J > 0, so that the solute will eventually demix from the solvent at a sufficiently low temperature and we will have precipitation. Kinetically, the solvent is considered to facilitate relaxation of the solute particle structure σ (i.e. the solvent is a plasticizer). In the model, this behaviour is modelled by associating a permanent up spin on each solvent. Finally, we have to allow solvent and solute exchange positions with a probability that is consistent with the constraints already imposed. Our basic rule is that solvent-solute neighbour pair exchanges in which either particle is in a site that would not allow that particles spin to flip, either before or after the exchange, are not permitted. Mathematically, a solvent particle on lattice site i and a solute particle (with spin $\sigma_j$, either 1 or 0) on neighbouring lattice site j will swap places with the following probability,

$$T_{ij} = H(m_i - 3)H(m_j + \sigma_j - 4)\min\{1, \exp(-J \times \Delta n_{ij} / T)\} \qquad (1)$$

where H(x) = 1 if x ≥ 0 and 0 if x < 0, $m_i$ is the number of up spins neighbour to site i and $\Delta n_{ij}$ is the change in solvent-solute neighbour pairs due to the swap. The requirement that the movement of the solute into or out of solution obeys the same cooperative kinetics as that which determines structural relaxation in the pure glass forming solute is core to the treatment of precipitation presented here and Eq. 1 represents the explicit statement of this core assumption. (The details of how we treat the relaxation of structure (i.e. the solute spin) in solution is a subtle question that we address in detail in the Supplementary Material.)

.

In Fig. 1b we present the T-x phase diagram for our model. The details of the calculation are provided in the Supplementary Material. The physics of amorphous precipitation is characterised by two temperatures: $T_c$, the critical temperature for demixing of the solution, and $T_g$, the glass transition of the bulk glass. These two temperatures are set, in turn, by two characteristic energies – the heat of mixing and the activation energy, respectively - J and h in our model. To get some idea of the relative magnitudes of these two energies we shall consider the specific case of a well-studied glass forming liquid, o-terphenyl (OTP). For OTP, the activation energy at high T is 1.03 kJ/mol [25] while the heat of mixing of OTP in benzene is 0.22 kJ/mol [26]. Translated to our lattice model, this means that if we set the activation energy h = 1.0, then we are interested in J ~ 0.21. In this study we have chosen J = 0.3 for most of the calculations presented but we shall consider the impact of varying this quantity later in the paper.

Having settled on the solute and solvent parameters, precipitation is controlled by the temperature and the initial concentration. Studies of vapor-deposited amorphous films [1-3] have established that the slower the rate of deposition, the more stable the amorphous state that is formed. The stability of the vapor deposited glasses also depends critically on the temperature of the substrate on which they are deposited with an energy minimum of the deposited film found at a substrate temperature roughly $0.8-0.9T_g$. In precipitation, the rate of aggregation is determined by the initial solute concentration $x$. The lower x, the lower the nucleation rate of precipitates clusters and the further each solute particle must travel, on average, before deposition. If we consider initiating precipitation by an instantaneous drop in temperature to some final value T, we can ask how the energy of the precipitate depends on the choice of T. In Fig. 2 we have plotted the energy per particle in the largest cluster formed as a function of the quench temperature T for a range of different initial solute concentrations. We have defined a cluster so as to exclude the surface contribution to the energy. To this end, a cluster consists of connected particles, solute or solvent, with 4 or more solute neighbours.

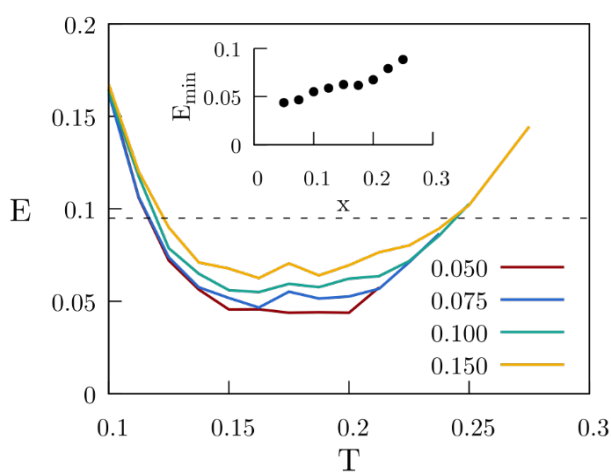

**Figure 2.** The final energy of the largest solute cluster (defined in the text) as a function of the quenched temperature T for a range of different initial solute concentrations. The final energy of the pure solute obtained by a cooling rate of $dT/dt = 10^{-12}$ is indicated by the dashed horizontal line. Inset: The value of the minimum energy as a function of the initial solute concentration.

The energetics of the precipitated clusters are presented in Figs. 2 and 3 which constitute the major results of the paper. In Fig. 2, we show that the energy of the amorphous precipitate exhibits a clear minimum with respect to $T$, analogous to the behaviour of vapor-deposited glasses [1-3]. This optimal quench temperature is roughly 0.4 of the bulk $T_g$ or ~ $0.7T^*$, with the latter relation roughly similar to the optimal T for vapor deposition (i.e. ~ $0.8T_g$). We also establish (see insert) that the value of this minimum energy decreases significantly as we decrease the initial solute concentration, again similar to the reported dependence of energy on deposition rate in from the vapor. In Fig.3a we compare the energy of the precipitated glass with the obtained by quenching a bulk sample with the same composition at a variety of cooling rates. We find that the precipitate energies can be well below that of the bulk glass and correspond to effective cooling rates over 7 orders of magnitude slower than that accessible to simulation. In addition to a low enthalpy, an ultra stable glass must exhibit enhanced kinetic stability [2]. In Fig. 3b we establish that the onset temperature at which the precipitated glass transforms into the supercooled liquid, on heating, is significantly higher than that exhibited by a sample at the same composition but formed via a temperature quench from above $T_g$. The kinetic stability of the precipitate formed at a solution concentration of x = 0.05 is similar to that of the (pure) vapor deposited glass for the same model [23].

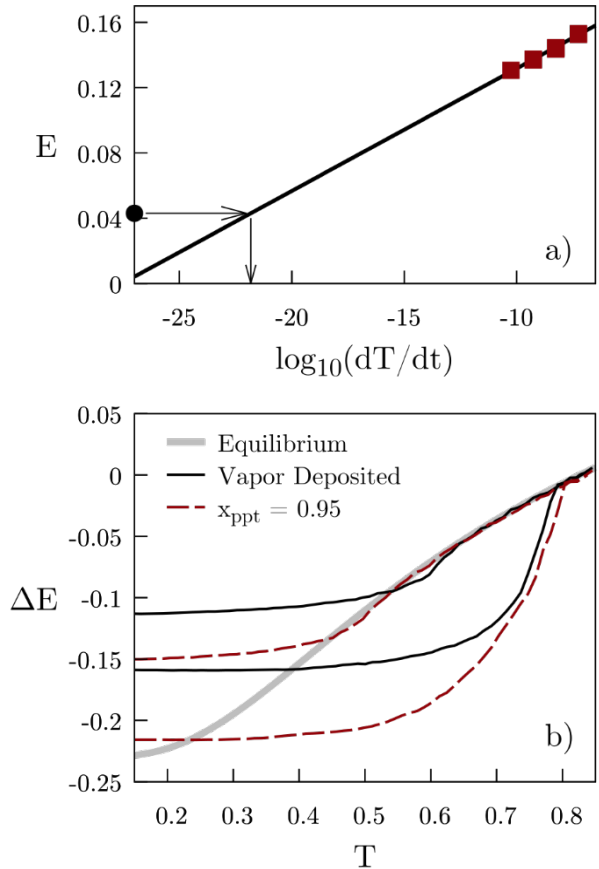

**Figure 3.** a) The energy of a bulk glass at the same composition (i.e. $x_{ppt} = 0.97$) as that of the precipitate (obtained from a solution with $x = 0.05$) as a function of the cooling rate used in its formation (red squares) is extrapolated to lower cooling rates (solid line). The minimum energy of the precipitate formed from an initial concentration of $x = 0.05$ is indicated by a filled circle on the energy axes with the effective cooling rates indicates by the arrow construction. b) The energy $\Delta E = E_{total}(T) - E_{total}(0.8)$ vs T during heating runs from stable deposited/precipitated glass (lower curve) and the as-quenched glass (upper curve) for precipitates with a concentration $x_{ppt} = 0.95$ (dashed line) and for the pure vapor deposited glass [23] (solid black line). The precipitate exhibits a similar kinetic stability as that found for the model vapor deposited glass.

The energy E of the precipitate can be resolved into two contributions: $E = E_{mix} + E_{struc}$ where $E_{mix}$ is the energy associated with mixing of solvent and solute, and $E_{struc}$ is the energy associated with the structural fluctuations of the amorphous solute (i.e. the spin energy in this model). All three energies are plotted in Fig. 4 for a quench to T= 0.17 for a range of initial solute concentrations $x$. We see that the energy E is non-monotonic in $x$. Starting from the pure solute case, the precipitate energy increases sharply as solvent is included, a direct consequence of the high energy of solvent-solute interactions as seen by the comparison of the variation of $E_{mix}$ and $E_{struc}$ over this concentration range. While the structural energy decreases towards its equilibrium value, thanks to the kinetic facilitation afforded by the plasticizer solvent, the energy associated with mixing solute and solvent, $E_{mix}$, increases. As $x$ decreases further, the energy E exhibits a maximum and then subsequent decrease. This decrease is entirely due to the decrease in $E_{mix}$, a result of the increasing segregation of solute from solvent as the slower growth of aggregation permits the precipitate to approach the (metastable) equilibrium concentration of the solute-rich phase as determined by the binodal line as shown in Fig. 2. Importantly, this increasing segregation of solute from solvent does not interfere with the enhanced relaxation of structure.

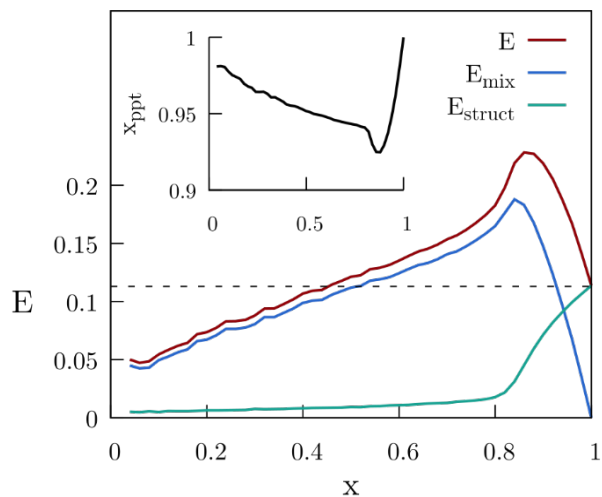

**Figure 4.** The energy E and its components, $E_{mix}$ and $E_{struc}$ of the amorphous precipitate as a function of the initial solute concentration x following an instantaneous quench from T = 0.5 to T = 0.17. The energy of the pure glass is indicated by the dashed horizontal line. Inset: the solute concentration $x_{ppt}$ of the precipitate as a function of the initial solution concentration *x*.

The picture, then, of the formation of ultra-stable glasses via precipitation from a plasticizing solvent is that structural relaxation, whose slow down ultimately determines the energy of the bulk glass, is kinetically enhanced to the point that the structural energy of the glass is reduced to close to its equilibrium value, something of a holy grail in glass physics. The cost of this kinetic facilitation is that the lowering of the structural energy is now compensated by the increase in energy associated with the positive heat of mixing of solute and solvent. We can summarise the general requirements for the solute/solvent system. These are: 1) the solute is a glass former, 2) the solvent $T_g$ lies well below that of the solute (i.e. the solvent will act as a plasticizer), 3) the solute glass transition temperature satisfies $T_{f,solvent} \ll T_g < T_{b,solvent}$, where the bounds are set by the freezing and boiling points of the solvent, respectively, and 4) $T^*$, as defined in Fig. 1a, lies well above $T_{f\,solvent}$. This last condition is the only one in which the solubility of the glass former in the solvent enters, even if implicitly. On this basis, we nominate the n-hexane/o-terphenyl system as a possible candidate, based on criteria 1) (o-terphenyl has a $T_g$ = 243K [26]), 2) (n-hexane has a $T_g$ = 70K [27]) and 3) (the melting and boiling points of n-hexane are 178K and 341.8K, respectively [28]).  The available data is insufficient to assess the final criterion.

In conclusion, we have demonstrated that a physically reasonable model of glassy relaxation kinetics can precipitate from a plasticizing solvent into ultra-stable amorphous solids with energies much lower than those kinetically accessible to bulk cooling. The essential 'recipe' for the formation of ultra-stable precipitates is to quench to a temperature below T$^*$ and to use as dilute a solution as practicable to minimise solvent trapping. The final energy of these very low energy precipitates is, we predict, dominated by the presence of residual solvent, rather than the intrinsic structural fluctuations of the glassy solute itself.

**Acknowledgements**

We acknowledge computational support from The University of Sydney High Performance Computing Service

**Supplementary Material**

**The Formation of Ultra-Stable Glasses via Precipitation: a Modelling Study**

Ian Douglass and Peter Harrowell

School of Chemistry, University of Sydney, Sydney NSW 2006 Australia

Contents:

## 1. The Calculation of the Solute-Solvent Liquid-Liquid Coexistence Curve

To determine the temperature dependence of the coexistence between solvent-rich and solute-rich solutions, as shown in Fig. 1b, we carry out simulations with the two phases in contact. The initial concentrations are selected to be close to the expected concentration in coexistence. Particle are allowed to exchange between the two solutions until equilibration is reached and the two concentrations recorded. The next step would be to slightly alter the temperature and, starting with the previously equilibrated coexisting solutions run a new equilibration run. The problem with this plan is that at low T we run into very slow equilibration times due to the glass transition. Instead, we keep T fixed (and well above $T_g$) and vary J instead since the equilibrium diagram only depends on the ratio J/T. Finally, to plot the diagram in the T-x plane (as opposed to the J/T-x plane) we select a value of $J^*$ and generate temperatures using T = $[T/J]J^*$. This is how the curves for different J's in Fig. 1b were generated. An example of the resulting phase diagram is shown in Fig. S1.

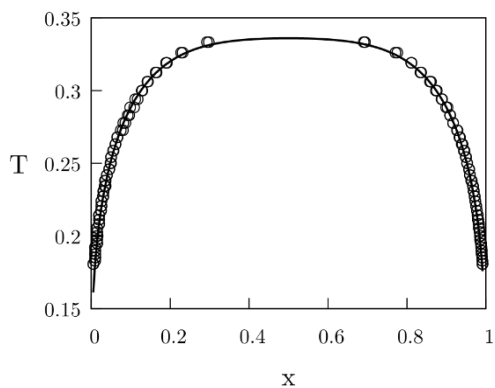

**Figure S1** The simulated coexistence concentrations of solvent-rich and solute-rich solutions plotted in the T-x plane, where x is the concentration of solute. A value of J = 0.3 was used to convert T/J to a temperature. The curve is the fitted function given in Eq. S1.

We have fitted the coexistence curve using the following function

$$T = \frac{k(x-0.5)^3 - 2T_c(1-2x)}{\ln x - \ln(1-x)} \qquad \text{(S1)}$$

where $T_c = 1.12J$ and $k = 5J$.

## 2. On the Treatment of Structural Relaxation of the Solute in Solution

One striking feature of the stable precipitates shown in Fig. 4 is the proximity of the structural energy $E_{struc}$ to its equilibrium value. Our model treats this structural relaxation very simply, assigning a spin to represent the two possible structural states of that solute particle. However useful that approach is in the bulk, how meaningful is it when the solute is in solution? Specifically, is the efficient relaxation of the structure as represented by these spins simply an artefact that we allow the spins of the solutes to flip, unimpeded, in solution where, perhaps, we should not be assigning the spin any physical relevance? To test the robustness of our results to this aspect of the model, we have adjusted the flip rule so that a solute spin can only flip if $n$ or more of its neighbours are also solutes. The calculations so far have implicitly used a value of n = 0. In Fig. S2 we plot the precipitate energy as a function of $0 \leq n \leq 6$. We find that E is independent on $n$ up to $n = 3$. This value of $n$ is sufficient to restrict spin flips to solute particles as they attach to the aggregate. The observation that the energy is largely unchanged from that where spin flips in solution are permitted (i.e. n = 0) indicates that the results presented here are unaltered if relaxation if the spin relaxation is restricted to the surface.

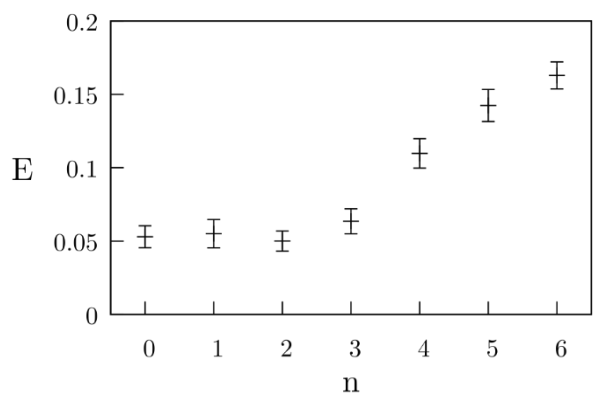

**Figure S2.** A plot of the precipitate energy as a function of $n$ for T = 0.17 and x = 0.1. As explained in the text, $n$ refers to the number of solute neighbours a solute must have before spin flips are permitted.

Fig. S2 raises a subtle issue associated with amorphous deposition. At what level of aggregation does it become meaningful to think about low and high energy solute-solute configurations? One possible answer is that the distinction can be made at the level of pairwise association (i.e. from n=1). This is certainly a plausible scenario in the case of complex molecules that would have number of pairwise arrangements to choose from. In the case of atomic aggregation, however, the pairwise aggregate has only one possible state and so we would have to wait for a larger aggregate in order to get a structural space of sufficient complexity to provide physically significant high and low energy possibilities. There is a trade-off between the minimal cluster size required to access low energy configurations and the maximum coordination at which any type of kinetic enhancement persists. Given the observation of ultra-stability in molecular films, we can conclude that these cases belong to the former case (i.e. stabilization occurring from small clusters) – as modelled here.

### 3. The Influence of J on the Solvent Concentration of the Precipitate

The calculations presented here have been carried out for a specific ratio $T_c/T_g = 0.72$. How does the relative size of these temperatures effect the stability of the amorphous precipitate? In Fig. S3 we have plotted the precipitate energy E as a function of the precipitation temperature T for different values of the heat of mixing J. We find that the minimum energy of the precipitate decreases with increasing J due to the improved segregation of the solvent from the solute. This effect depends on the use here of an instantaneous quench to the final T. If a linear cooling rate is used, the higher J (and hence higher $T_c$) will result in precipitation starting at a higher T and, hence, an increase in the energy of the precipitate. We conclude that the influence of J (or $T_c$) on the stability of the precipitate will be complex, arising from the competition of a number of factors.

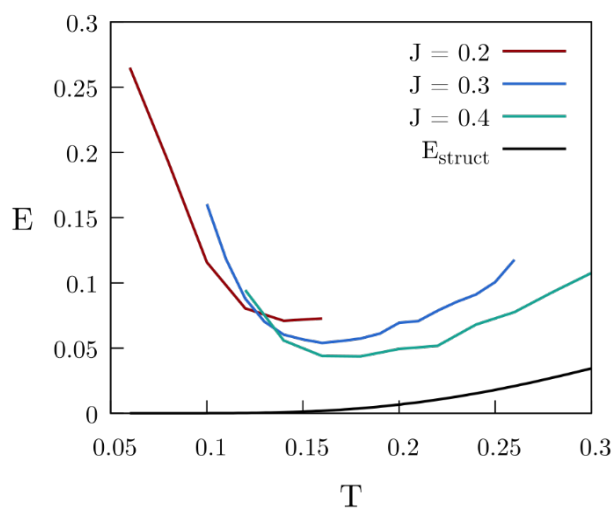

**Figure S3.** The dependence of the precipitate energy E, along with $E_{struct}$, as a function of T for various J. The precipitation involved a quench from T = 0.5 with an initial concentration of x = 0.1. We see a shift in the temperature of the minimum energy, as well as an increase in E for smaller J.

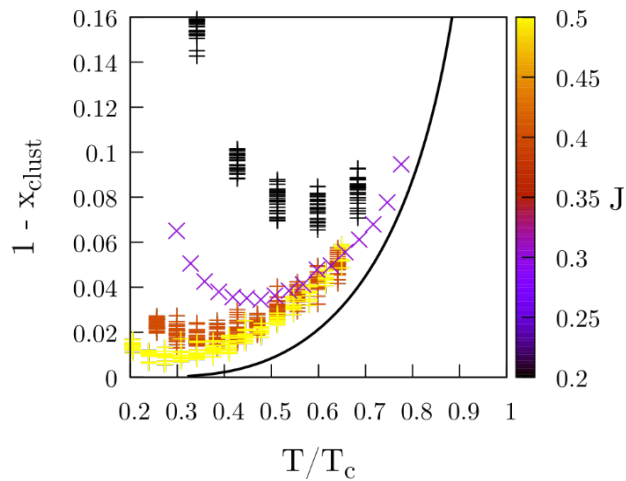

**Figure S4.** The solvent concentration, 1-$x_{cluster}$, in the precipitates as a function of $T/T_c$ for different values of J. The values of J are indicated by the right hand color scale. The equilibrium solvent concentration is shown as the solid line.

In Fig. S3 in the text we showed that the minimum energy of precipitate decreases with increasing heat of mixing J. This rather counter-intuitive result is a consequence of the improved segregation of solute and solvent with increasing J. To see this, we have plotted in Fig. S4 the concentration of solvent, 1-$x_{cluster}$, as a function of T for different values of J. By plotting the temperature as $T/T_c$, the equilibrium solvent concentration (solid curve) is independent of the choice of J. We see in Fig. S2 that as J increases (i.e. symbol colors changing from black to yellow, the solvent concentration drops towards the equilibrium value. The increase in precipitate energy as we increase T is clearly shown in Fig. S3 to be a track with the increasing equilibrium solvent concentration – a rigorous lower bound on the amount of solvent remaining in the precipitate.

## 4. Structure of Precipitated Clusters

In Fig. S5 we show images of the final precipitates formed at temperatures above and below the optimal temperature. The optimally stable precipitate forms a single compact solid with a minimum residual solvent concentration. In contrast, low T precipitate exhibits a highly ramified (non-compact) morphology characteristic of aggregation without relaxation. Precipitates at a temperature above the optimal temperature are compact but with a larger solvent inclusion and a rougher surface.

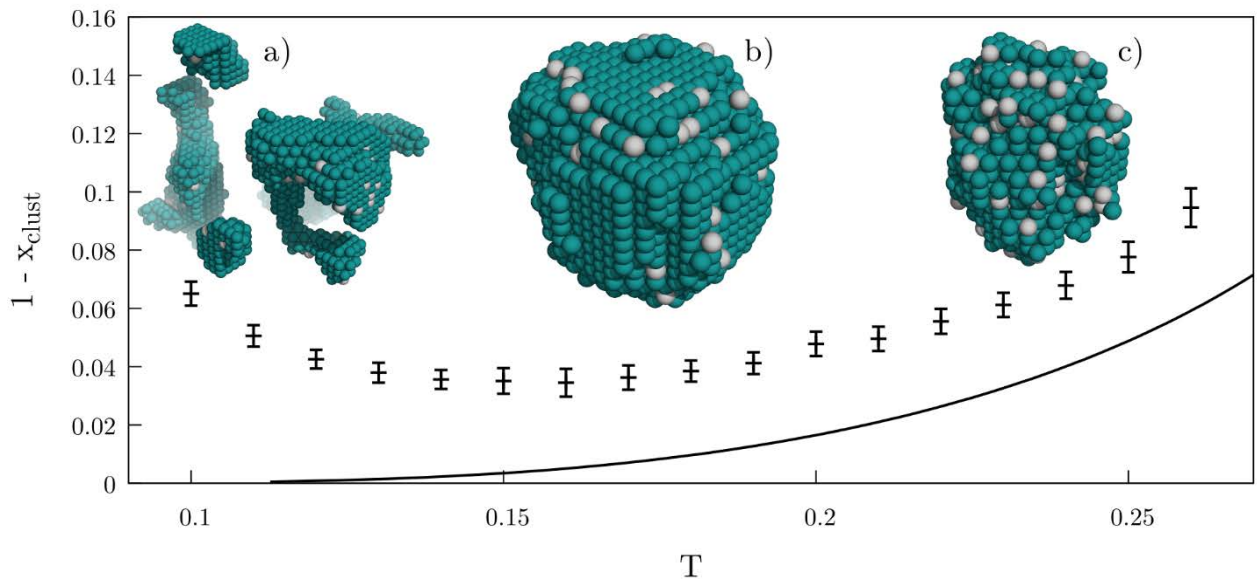

**Figure S5.** Examples of the final solid aggregate morphology for a precipitation at a) T = 0.1, b) T = 0.15 and c) T = 0.25, all from an initial concentration of x = 0.1. The solute particles are colored blue and the solvent grey. Also shown is the solvent concentration 1-$x_{cluster}$ as found in the precipitate (symbols) and the equilibrium value of the solvent concentration (solid line).

## 5. Precipitation of an Ultra-Stable Glass onto a Planar Substrate

To simulate precipitation on to a planar substrate, use a simulation cell periodic in the X and Y axes, but not in the Z. We construct a substrate normal to the Z axis consisting of a pure equilibrated (solute) glass and do not allow spin flipping in or solvent exchange into this layer, i.e. we pin the structure of the substrate. In order to maintain a roughly constant concentration of solute in solution during precipitation, we have constructed a layer of pure solvent, parallel to the substrate and on the opposite side of the simulation cell. Additional solute is added the system as follows. Every 500 cycles we replace the five layers below the pure solvent layer with a randomly generated solution with the specified glass concentration. At all other times the simulation progresses as previously detailed. Energy is calculated for the cluster including the glass layer, and layers as number of precipitated particles (including solvent) in this cluster divided by the area of the system.

In Fig. S6 we plot the growth, as measured by number of layers, of the precipitate on the substrate. We also plot the total energy. Our optimally stable glass, as shown in Fig.2, has an energy $E_{total} = 0.05$. In Fig. S6 we show that a substrate precipitated glass achieves a similar stability, with $E_{total} \sim 0.05\text{-}0.06$

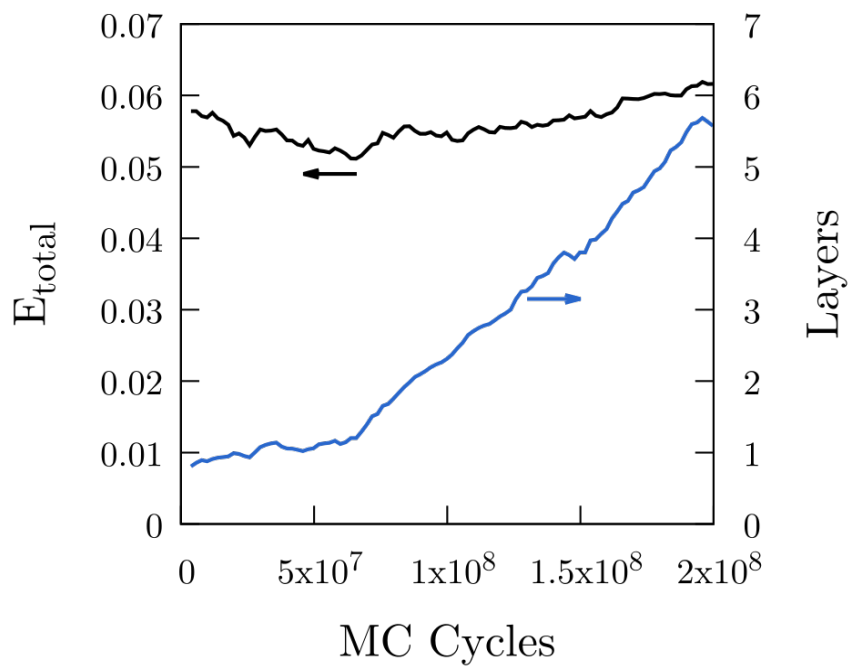

**Figure S6.** A plot of the total energy E<sub>total</sub> and the thickness of the precipitated films (in number of layers). This calculation was carried out at T = 0.172 with a solute concentration of 0.067.